\documentclass[twocolumn,showpacs,preprintnumbers,amsmath,amssymb]{revtex4-1}
\usepackage[english]{babel}
\usepackage{ucs}
\usepackage[utf8x]{inputenc}
\usepackage[dvips]{graphicx}

\begin{document}

\title{Atomistic simulations of the implantation of low energy boron
  and nitrogen ions into graphene}

\author{E.~H. {\AA}hlgren$^{1}$, J. Kotakoski$^{1}$ and A.~V. Krasheninnikov,$^{1,2}$}

\affiliation{$^1$ Department of Physics, University of Helsinki,
  P.O. Box 43, 00014 Helsinki, Finland\\ \\ $^{2}$ Department of
  Applied Physics, Aalto University, P.O. Box 1100, 00076 Helsinki,
  Finland}

\date{\today}

  \begin{abstract}

    By combining classical molecular dynamics simulations and density
    functional theory total energy calculations, we study the
    possibility of doping graphene with B/N atoms using low-energy ion
    irradiation. Our simulations show that the optimum irradiation
    energy is 50~eV with substitution probabilities of 55\% for N and
    40\% for B. We further estimate probabilities for different defect
    configurations to appear under B/N ion irradiation. We analyze the
    processes responsible for defect production and report an
    effective swift chemical sputtering mechanism for N irradiation at
    low energies ($\sim$125 eV) which leads to production of single
    vacancies. Our results show that ion irradiation is a promising
    method for creating hybrid C-B/N structures for future
    applications in the realm of nanoelectronics.

  \end{abstract}


\maketitle

\section{Introduction}

Graphene~\cite{Novoselov2004} is a two-dimensional all-carbon
nanostructure consisting of $sp^2$-bonded carbon atoms arranged in a
honeycomb-like lattice.  Similar to carbon
nanotubes~\cite{Iijima1991}, which can be imagined as graphene sheets
rolled up to form seamless tubes, it has attracted a considerable
amount of attention, especially due to its unique electronic
properties~\cite{CastroNeto2009}. Pristine graphene is a zero band gap
material with a high electron mobility, but introduction of impurities
and structural defects~\cite{CastroNeto2009,Appelhans2010} can lead to
band gap opening and Fermi level shifting. This makes graphene an
interesting candidate for many applications in nanoelectronics.
However, despite recent breakthroughs in graphene
production~\cite{Li05062009,Bae10nn}, one of the remaining problems on
the way towards graphene-based electronic devices is the lack of
control over the electronic properties of the as-grown samples.

In today's semiconductor industry, a standard way to modify the
electronic properties of a material is doping with foreign atom
species. For an all-carbon material, such as graphene, the natural
dopants are boron and nitrogen, which posses one electron less and
more than carbon, respectively. Since they are also roughly similar in
size to carbon, they can be incorporated into substitutional positions
in carbon structures by replacing exactly one carbon atom, or take the
vacant site at the edges of graphene
flakes~\cite{Yan2007,Huang2007,Reddy10acs,Wang09sci,Wei09nlD}. It has
been shown~\cite{Late2010} than B/N-doped graphene displays
$p$/$n$-type behavior. Not surprisingly, several theoretical studies
on hybrid C-B/N
systems~\cite{Zheng2008,Lherbier2008,Peres2007,Zheng2010a,Yu2010b,Wang2010b}
have recently been carried out. Particular attention has been paid to
graphene
nanoribbons~\cite{Zheng2009a,Pekoz2009,Biel2009,Yu2010,Yu2010a,Huang2010}.
Although it is possible to introduce B/N atoms into graphene during
synthesis~\cite{Reddy10acs,Wang09sci,Wei09nlD,Ozaki2007,Lin2009,Ci2010,Raidongia2010},
not many studies have investigated the possibility of selectively
introducing B/N impurities into graphene after growth, while
post-synthesis doping may be an alternative way to create
graphene-based functional materials. It was demonstrated in a recent
study that B can be incorporated into graphene structure by
selectively exposing different sides of a graphene monolayer to
different atomic environments~\cite{Pontes2009}. Unfortunately, it is
not clear how this could be carried out in practice. While ion
irradiation is routinely used nowadays when manufacturing conventional
semi-conductor devices, one can expect that it can be used in a
similar manner to introduce B/N atoms into graphene. However,
graphene, an atomically thin target, has a significantly different
response to ion irradiation as compared to traditional
three-dimensional materials~\cite{Lehtinen2010}. In a recent
experimental study~\cite{Guo2010}, N$^+$ ion irradiation was used to
facilitate N-doping of graphene. Due to a high irradiation energy
(30~keV), very few if any ions were directly incorporated to the
graphene sheet. Instead, the nitrogen atoms were later introduced
during annealing in a NH$_3$ atmosphere. Clearly, the optimization of
the implantation process requires full microscopic understanding of
the response of graphene to ion irradiation.

In this study, we employ classical molecular dynamics (MD) simulations
and density functional theory (DFT) total energy calculations to study
the efficiency of ion beam doping of graphene as a function of ion
energy. We show that at energies close to 50~eV, it is possible to
incorporate as much as 55\% of N and almost 40\% of B impinging ions
to the substitutional dopant site in graphene. The most typical
defects which appear under low energy irradiation include single and
double vacancies. In the preferred range of doping energies, the
probability for introducing additional defects to graphene during ion
beam doping remains below 37\%.  We also report a swift chemical
sputtering mechanism (especially for N ions) for ion energies below
150~eV, which along with the ballistic knock-out, leads to production
of single vacancies. This effect may have an important role in defect
creation during experimental ion beam doping of graphene, and possibly
other carbon materials.

\section{Methods}

To study B/N ion implantation into graphene, we followed the same
approach as in our earlier work on ion irradiation of carbon
nanotubes~\cite{Kotakoski2005,Kotakoski2005a,Tolvanen2007,Kra05chan} and
graphene~\cite{Lehtinen2010}. The interactions between the atoms in
the system under study were described by the bond order potentials
developed by Matsunaga {\it et al.}~\cite{Matsunaga1999} for C-B/N and
by Brenner {\it et al.}~\cite{Brenner1990} for C-C interactions
(without the bond conjugation term). Heat dissipation at the
boundaries of the system was modeled with the Berendsen
thermostat~\cite{Berendsen1984} with various time constants.  However,
for the used system size and rather low irradiation energies (up to
4~keV) included in this study, the choice of the time constant had
only a minor effect on the results. Our graphene sheet consisted of a
$16\times 16$ supercell with a total of 512 atoms. During the
simulations, we directed the B/N ion towards graphene in perpendicular
direction with energies between 10~eV and 4~keV. 1000 impact points
were selected randomly within the irreducible area of the graphene
lattice to ensure statistically correct sampling of the irradiated
structure. The total number of independent simulations was nearly 1
000 000.

We repeated the simulations with the Tersoff C-C
potential~\cite{Tersoff1986} to get an insight on how sensitive the
results are with regard to the choice of the potential.  No
significant differences were observed. Moreover, since the Brenner
potential gives a displacement threshold energy (minimum kinetic
energy required to sputter a carbon atom) which is very close to that
obtained with dynamical DFT calculations (22.20~eV as compared to
22.03~eV~\cite{Kotakoski2010}), we are confident that the defect
production upon irradiation is at least qualitatively correctly
described within our simulation model. The same interaction models
were also used in our earlier simulations regarding ion implantation
into carbon nanotubes~\cite{Kotakoski2005a,Kotakoski2005}. For these 
systems, our theoretical predictions on 
nitrogen implantation have been later 
experimentally corroborated~\cite{Morant2006}. The charge of the ion was not
explicitly taken into account in our simulations, since it is known to
have only a minor role in the defect production during ion irradiation
of carbon nanostructures~\cite{Krasheninnikov2010}. However, we refer
to the projectile as an ``ion'' to facilitate the direct juxtaposition
with the experimental studies.

As the C-B/N interactions in our model have been tested much less 
than the C-C interaction, we employed DFT total energy calculations to ensure that the
resulting defect structures involving B/N atoms, as given by the MD
simulations, are consistent with {\it ab initio} results. Our DFT
calculations were carried out with the {\scshape VASP} simulation
package~\cite{Kresse1996a, Kresse1996} using projector augmented wave
potentials~\cite{Blochl1994} to describe core electrons, and the
generalized gradient approximation~\cite{Perdew1996} for exchange and
correlation. Kinetic energy cutoff for the plane waves was set to
500~eV, and all structures were relaxed until atomic forces were below
0.01~eV/{\AA{}}. Our initial structure consisted of 200 carbon atoms,
so that the finite size effects are expected to be small. The Brillouin
zone sampling scheme of Monkhorst-Pack~\cite{Monkhorst1976} with up to
$5\times5\times1$ mesh was used to generate the $k$-points.

\section{Results and Discussion}

\begin{figure}
\includegraphics[width=.45\textwidth]{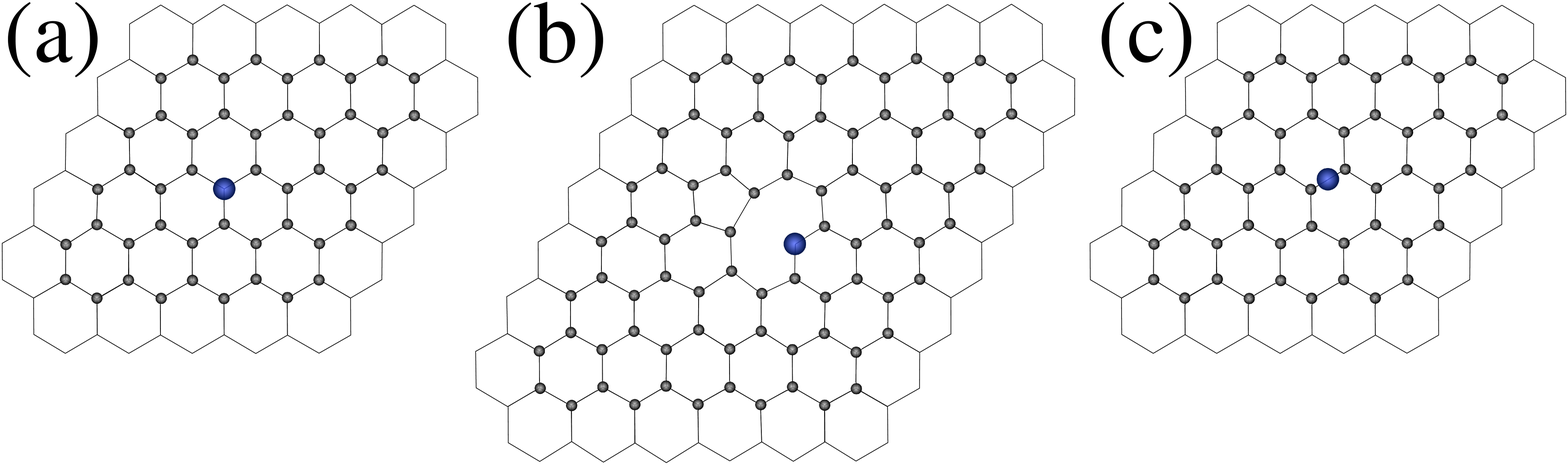}
\caption{(Color online) Typical defect structures in graphene 
 observed after low-energy irradiation with B and N ions, 
 as calculated with MD method, followed by structure
relaxation using DFT calculations: (a) perfect
  substitution of one carbon atom with the dopant, (b)
  dopant-vacancy--complex structure in which the dopant atom is bonded
  to two carbon atoms, and (c) dopant atom in the bridge-configuration
  as an adatom on top of a perfect graphene lattice.}
\label{fg::kuva1}
\end{figure}

\begin{figure}
\includegraphics[width=.45\textwidth]{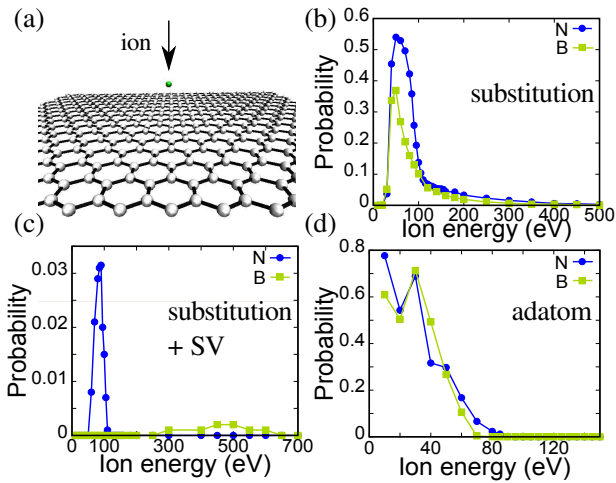}
\caption{(Color online) (a) Schematic representation of the simulation
  setup. Probabilities for different defect configurations as
  functions of the ion energy: (b) perfect substitution, (c)
  dopant-vacancy--complex structure, (d) dopant as an adatom on top of 
the graphene sheet.}
\label{fg::kuva2}
\end{figure}

We started our simulations by calculating the probabilities for an
irradiation event to produce the most typical defect structures (some
of which are shown in Fig.~\ref{fg::kuva1}), as functions of ion
energy for both B and N ions. The most prolific outcomes were
substitution (exactly one C replaced by the B/N), substitution with a
neighboring vacancy, single vacancy (SV) and double vacancy (DV). At
low energies, we also frequently observed the dopant atom attached to
the perfect graphene structure as an adatom. A schematic of the
simulation setup as well as the defect creation probabilities of these
structures are presented in Fig.~\ref{fg::kuva2} for those defects
involving the dopant atom (presented in Fig.~\ref{fg::kuva1}), in
Fig.~\ref{fg::kuva3} for SV and in Fig.~\ref{fg::kuvaDV} for DV.

\subsection{Incorporation of the dopant atom into graphene}

The highest observed probability (55\%) for any of the structures
(except for adatom deposition at very low energies which naturally
gives a probability of 100\%) was seen for N substitution
(Fig.~\ref{fg::kuva1}b) at energies close to 50~eV
(Fig.~\ref{fg::kuva2}b). For substitution, it is intuitively clear
that the probability should display a single peak at these energies:
although the displacement threshold for a carbon atom is about 20~eV
in a head-on collision, on average the ion must have a somewhat higher
energy in order to displace the atom, as a nearly head-on collision is
unlikely. For B substitution the maximum is slightly below 40\%
(Fig.~\ref{fg::kuva2}b), and it appears at a slightly lower energy
than for N. The lower energy for B ion is a results of the difference
between the momentum transfer from B/N ions to the target atom while
the lower probability arises from the smaller displacement cross
section for the B ion.  By comparing to noble gas ion irradiation
results for single vacancy production in graphene~\cite{Lehtinen2010}
(substitution occurs when the ion is stopped after a successful
sputtering of one target atom), it is easy to see that these features
are indeed caused by the mass difference between the ions rather than
chemical effects. For energies higher than 0.5~keV, very few ions were
incorporated into the graphene atomic network. Instead, at high
energies (where chemical effects are not important, as shown below)
relatively larger vacancy-type defects will be
produced~\cite{Lehtinen2010}.

For nanotubes~\cite{Kotakoski2005,Kotakoski2005a}, the highest
observed probability for B and N substitutions were similar to those
found for graphene, although the exact values were slightly
lower/higher for N/B. For both ions, the peaks in graphene appear at
lower energies and are significantly narrower. This can be understood
based on the geometrical differences between nanotubes and
graphene. As a crude approximation, one can think of the nanotube as a
parallelepiped with two walls perpendicular to the ion beam direction
and two walls parallel to it. Therefore, it is possible for the ion to
be slowed down by the first wall before the displacement takes
place. This naturally leads to substitution events also at energies
higher than what is observed for graphene, a two-dimensional
system. What is perhaps more surprising is that the probabilities
themselves are not much different. {\it A priori} one could have
expected the nanotube structure to yield roughly twice the number of
substitutions due to the two walls parallel to the beam. However,
since it also takes energy to penetrate the first wall, this is
manifested only at energies above those for the peaks observed in
graphene.

Another type of observed substitutions is a result of an irradiation
event in which the ion has first displaced two carbon atoms from
graphene and then remains in the place of one of the displaced atoms
in a pyridine-like configuration \cite{Ewels05rev}, as shown in
Fig.\ref{fg::kuva1}(b).  We never observed pyrrolene-like structures
\cite{Ewels05rev} in which the nitrogen/boron atom occupies a site in
a five-membered atomic ring. This situation may be more probable in
nanotubes~\cite{Lin2009} where such configurations are assisted by the
curvature. Since it was not seen in the nanotube ion irradiation
simulations~\cite{Kotakoski2005,Kotakoski2005a} either, it may also be
a feature which only occurs when nitrogen dopants are introduced
during synthesis.

Since two atoms must be displaced for a dopant-vacancy--complex to be
created, it correspondingly occurs at higher energies than the perfect
substitution. The probability for this process is presented in
Fig.~\ref{fg::kuva2}c as a function of the ion energy. Surprisingly,
the probabilities are significantly different for B and N ions, with a
probability for N being larger by more than two orders of magnitude.
However, even in this case the probability maximum is only $\sim
3$\%. Again, the difference is mostly due to the mass difference
between the B and N ions. For the heavier N, with increasing energies
the scattering process leads to displacements of the carbon atoms in
the directions perpendicular to the initial ion direction, as
described in Ref.~\cite{Lehtinen2010}, while the ion only slightly
deviates from the original direction. Hence, the defect is produced by
simultaneous displacement of two carbon atoms while the ion itself
stops at the graphene sheet. However, for B, which is lighter than C,
the situation is opposite; with higher energies the ion itself is
scattered in a direction parallel to the plane where it will be
scattered again by another C atom and thus eventually leaves the
structure.

The probability for creating the dopant-vacancy--complex in
nanotubes~\cite{Kotakoski2005,Kotakoski2005a} was similar to graphene
values for N but significantly higher for B (at optimal
energies). This is due to the fact that the projected atomic density
in the direction parallel to the initial ion trajectory is in some
areas much higher for nanotubes than for graphene due to
curvature. Therefore, even though the displacement cross section for
the B ion with respect to target atoms is relatively small at energies
required to displace two atoms, simultaneous displacement of an atom
pair is still more likely than in the case of graphene.

As mentioned above, we checked all observed final structures with DFT
calculations to ensure their stability beyond the description provided
by the bond-order potential. Indeed, both of the substitution cases
remained stable during the structure optimization (the DFT-optimized
structures are shown in Fig.~\ref{fg::kuva1}). The formation energy
for the perfect substitution is ca. $\sim 4.42$~eV lower than that for
the pyridine-like two-coordinated configuration. Therefore, one can
expect that in the case of migrating carbon adatoms on the graphene
surface, the pyridine-like configurations tend to get filled with
carbon atoms to form a structure similar to perfect substitution. As
has been shown previously~\cite{Kotakoski2005a}, the formation
energies of the substitutional impurities in graphene are a few
electron-volts with a somewhat higher formation energy for boron
substitution.

\subsection{Creation of single vacancies }

\begin{figure}
 \includegraphics[width=.45\textwidth]{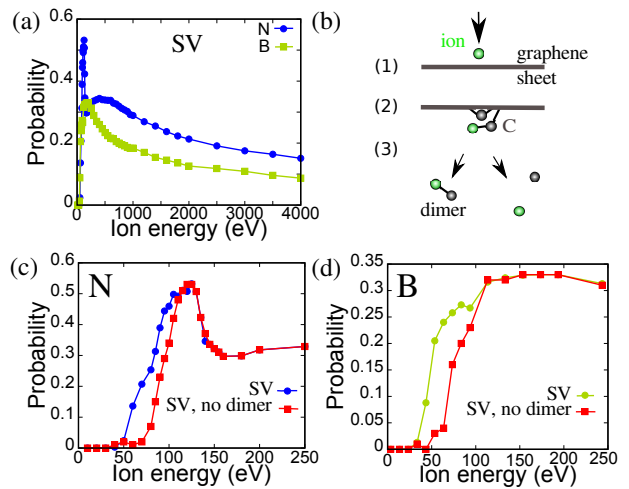}
 \caption{(Color online) Probabilities for creating single vacancies
   in graphene under low-energy B and N irradiation (a).  Schematic
   presentation of the swift chemical sputtering process involving
   chemical interaction between the incoming ion and the recoil carbon
   atom is shown in panel b. In panels c and d these probabilities are
   shown for N and B ions (respectively) with and without those cases
   in which a dimer is formed between the incoming ion and the
   sputtered carbon atom.}
 \label{fg::kuva3}
\end{figure}

In addition to substitutional impurities, ion-implantation should
inevitably result in the formation of various irradiation-induced
defects in graphene~\cite{Banhart2010,Krasheninnikov2010}.  Among
these, single vacancies have the highest probability to appear.  In
the case of N irradiation, it reaches its maximum at ca. 125~eV with a
value of 55\% (Fig.~\ref{fg::kuva3}a). This very narrow peak overlaps
with a broad peak with its maximum (35\%) at ca. 400~eV.
Surprisingly, for B, only a slight effect of the first peak is seen as
a shoulder of the second peak. In this case the peak has a maximum of
35\% at 180~eV (Fig.~\ref{fg::kuva3}a). The narrower low-energy peak
results from a chemical interaction between the ion and one of the
target carbon atoms. From a case-by-case analysis of the distance
between the ion and the sputtered atom as a function of time, it is
clear that the attractive interaction between the projectile and
recoil atom is responsible for the production of single vacancies at
energies within this peak via a mechanism where the ion pulls one
carbon atom with itself while penetrating the sheet, as sketched in
Fig.~\ref{fg::kuva3}b. This effectively results in the formation of a
dimer, which may eventually be broken, as also sketched in
Fig.~\ref{fg::kuva3}b.

In Fig.~\ref{fg::kuva3}c,d we show the effect of the attraction by
subtracting those cases where the ion and a carbon atom form a dimer
during the sputtering process from the overall probability for the
single vacancy creation. For B, this process completely accounts for
the shoulder of the single vacancy probability peak. However, for N,
it only accounts for the lower energy end of the narrow peak. This
effect can also be directly explained by comparing the masses of the
ions, and taking into account that both N and B have an attractive
interaction with carbon atoms. Because B is lighter than C, it is
unlikely that it will escape the carbon atom after successfully
displacing it at the energy range where the attraction plays a
role. Thus, the dimer remains stable after the sputtering
event. However, for the heavier N, the ion can effectively pull one
carbon atom after itself but still retain enough kinetic energy to
escape the atom thus breaking the dimer. By carefully comparing cases
with exactly the same impact parameters, except for a slight change in
kinetic energy of the ion, in which the lower energy produced a single
vacancy and the higher energy did not, we concluded that this swift
chemical sputtering process (in principle similar to what has been
observed in Ref.~\cite{Salonen2001} for hydrogen ions) indeed
completely explains the observed narrow peak in the single vacancy
production probabilities for both N and B ions.

These results are in line with Ar irradiation simulations carried out
for nanotubes~\cite{Tolvanen2007}, as well as with the noble gas
irradiation studies for graphene~\cite{Lehtinen2010}. If one ignores
the chemical sputtering which does not occur for the noble gases, the
single vacancy probabilities for B/N appear at energies between those
for He and Ne in the order of increasing mass, as one would
expect. Also the functional form is similar to that observed for He
and Ne irradiation, even reproducing the broadening of the probability
peak with increasing mass of the
projectile~\cite{Lehtinen2010}. Therefore, also this comparison shows
that the additional narrow peak must be caused by a chemical effect.

\subsection{Creation of double vacancies}

\begin{figure}
 \includegraphics[width=.45\textwidth]{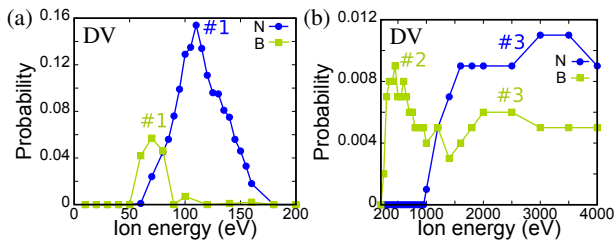}
 \caption{(Color online) Probabilities for creating double vacancies
   in graphene under impacts of B and N ions. The low-energy limit (a)
   and the probabilities at energies exceeding 200 eV. Note the
   difference in the scale of the $y$ axis. The numbers in the plot
   indicate the peaks as discussed in the text.}
 \label{fg::kuvaDV}
\end{figure}

\begin{figure}
\includegraphics[width=.45\textwidth]{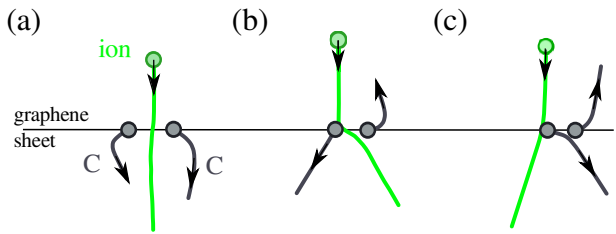}
\caption{(Color online) Example trajectories for the different double
   vacancy creation mechanisms. (a) A direct impact by the ion with
   two carbon atoms. The ion strikes in the middle of the bond joining
   these atoms and causes the sputtering of both.  (b) B ion, which is
   lighter than C, gets scattered in the in-plane direction of graphene
   so that it has enough kinetic energy to cause sputtering of a
   second carbon atom.  (c) Ion displaces one carbon atom which then
   causes another carbon atom to sputter.}
\label{fg::kuva4}
\end{figure}

The last of the often observed defects is DV. The highest probability
(16\%) for this defect was found upon N irradiation at energies close
to 110~eV (Fig.~\ref{fg::kuvaDV}a). For B, the corresponding peak
maximum occurred at 70~eV with a value of 6\%
(Fig.~\ref{fg::kuvaDV}b). In addition to this first peak, the
probability for B irradiation contains two other, much broader,
peaks. The first of these (\#2) occurs with a probability of less than
1\% at energies between 0.14~keV and 1.4~keV. The second one (\#3)
appears with a similar probability maximum at energies closer to
2~keV. This peak extends to energies higher than those included in
this study. For N, only the second one of these high energy peaks
(\#3) appears. From the three peaks, each of which occur at the
characteristic energy range, it's easy to predict that there are three
mechanisms for the double vacancy production. These three mechanisms
are illustrated in Fig.~\ref{fg::kuva4}.

The lowest energy peak is due to a direct impact by the ion with two
carbon atoms when the ion strikes in the middle of the bond between
these atoms (Fig.~\ref{fg::kuva4}a).  For energies exceeding the value
corresponding to the maximum in the curve, the scattering cross
section for the ion and a target atom becomes so small that it is
impossible for the ion to displace two atoms simultaneously. For the
heavier N ion, the correspondingly larger cross section extends the
probability peak over the energy range observed for B. Due to the
larger displacement cross section also the peak maximum appears with a
higher value for N. This also explains the difference observed for the
ions in the case of vacancy-dopant--complexes, which are directly
linked to double vacancies.

The second mechanism (Fig.~\ref{fg::kuva4}b), which does not occur for
N, arises from the mass difference between the B ion and carbon
atoms. Because B is lighter than C, at the intermediate energies it
gets scattered in the in-plane direction so that it has enough kinetic
energy left to sputter a second carbon atom. In the case of boron,
this mechanism leads to the highest probability for the
vacancy-dopant--complex defect in the case where the ion has enough
energy to displace both carbon atoms but becomes trapped in the carbon
network. For the heavier N, this never occurs since it is rather the
carbon atom which gets scattered in the in-plane direction. Similar
behavior is also evident in the noble gas irradiation
results~\cite{Lehtinen2010} where two peaks are observed for the Ne
ion irradiation (Ne mass is about 1.4 times larger than that of N),
similar to the N irradiation in the present study.

The third of the observed double vacancy creation mechanism
(Fig.~\ref{fg::kuva4}c) involves the displacement of one carbon atom
in the in-plane direction by the ion, which then causes the sputtering
of a second atom. In the case of noble gas ions heavier than Ne, the
two peaks (corresponding to peaks one and three for B irradiation)
overlap so that the separation of the peaks becomes
impossible~\cite{Lehtinen2010}.

\subsection{Dopant as an adatom}

Obviously, for ion energies below the threshold for creating a single
vacancy, the dopant will be attached to the pristine graphene
structure as an adatom (if it does not go through the middle of the
hexagon). Depending on the impact point and the energy of the
projectile, it is also possible that the ion gets bounced back from
the graphene sheet, which leads to a non-monotonous probability curve
as a function of the ion energy (Fig.~\ref{fg::kuva2}d). Although it
is rather unlikely that the adatom configuration would spontaneously
transform to a substitutional defect, it is possible that the adatoms
will take the position of a missing atom in a single or double vacancy
created during the irradiation, since the migration barriers for B/N
adatoms are in the range of 0.1--1.1~eV on graphenic
surfaces~\cite{Lehtinen2003,Kotakoski2005a}.

From a simple formation energy consideration at the thermodynamical
equilibrium, it would also be beneficial for nitrogen dimers to break
in order to fill in single vacancies in graphene (according to DFT
calculations, this would lead to a significant energy gain of more
than 6~eV per nitrogen atom). This could occur during annealing in a
N$_2$ atmosphere. However, this process would require the breakage of
the N dimer which is among the strongest molecules in nature. 
Recent experiments indeed
showed~\cite{Guo2010} that annealing in a N$_2$ atmosphere does not
lead to N-doping of defective graphene. However, using a molecular
environment in which the nitrogen (or boron) atoms are less strongly
bound, such as NH$_3$, will provide dopant atoms to fill the
vacancies, as was demonstrated in Ref.~\cite{Guo2010}.

\subsection{Relative probability for substitution}

\begin{figure}
 \includegraphics[width=.35\textwidth]{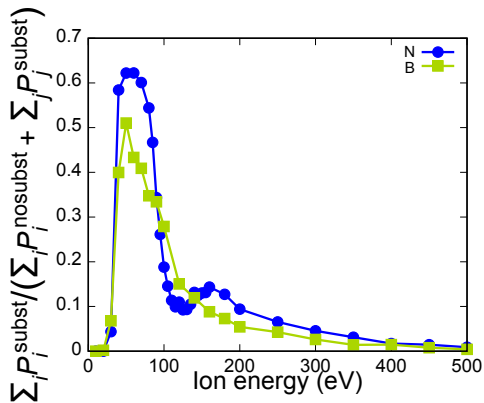}
 \caption{(Color online) Relative probability for B/N substitution as
    compared to creation of any other defects as a function of ion
    energy.}
 \label{fg::kuva5}
\end{figure}

Naturally, one of the critical factors in using ion irradiation to
dope a nano-object such as graphene, with plenty of free space for the
target atoms to escape, is the amount of other defects created during
the implantation. Luckily, the production of single vacancies, and
especially more complex defects, occurs mainly at energies higher than
that of the highest substitution probability. The total probability
for creating any other defect during nitrogen ion irradiation at 40~eV
is 29\%. For boron, the corresponding value at 100~eV is 27\%. These
energies correspond to 10\% substitution probabilities at the lower
and higher end for nitrogen and boron, respectively. At the most
efficient doping energy (50~eV) for N, the probability for creating
any other defects is 32\%. For B at this energy, in 36\% of the events
are other defects created. These relative probabilities, defined as
the ratio of sum of probabilities for the different substitution cases
($\Sigma_iP_i^\mathrm{subst}$) to the sum of probabilities for all
defects ($\Sigma_iP_i^\mathrm{nosubst} + \Sigma_iP_i^\mathrm{subst}$),
for B/N substitution are shown in Fig.~\ref{fg::kuva5} as a function
of the irradiation energy. An obvious way to overcome any problems
arising from creating a substantial amount of single vacancies during
the substitution is to combine irradiation with two energies; one at
the substitution maximum and another one below the single vacancy
creation threshold. This would allow deposition of dopant adatoms
which could fill in the created vacancies. Although we are unaware of
any experiments where such a method would have been used, this is in
many ways similar to that of annealing in a suitable molecular
atmosphere providing the dopant atoms, as presented in
Ref.~\cite{Guo2010}.

\section{Conclusions}

As we have shown above, ion irradiation can be used for implanting B/N
ions into graphene. The most efficient doping energies are, according
to our simulations, close to 50~eV for both ions. The high-probability
peaks are wide enough to allow doping also at various energies both
below and above the maximum. For example, substitution probability for
nitrogen ion remains over 10\% for energies between 40~eV and 105~eV,
and for boron between 40~eV and 100~eV. Assuming a control over ion
energies and the possibility to use focused ion beams, one can
envisage fabrication of spatially-non-uniform graphene based materials
where N and B doped areas can be made with a resolution of better than
5~nm. As the defect structures which we observed during our MD
simulations were confirmed to be at least metastable by our DFT
calculations, and because our earlier theoretical
predictions~\cite{Kotakoski2005, Kotakoski2005a} on ion implantation
into carbon nanotubes have been experimentally
corroborated~\cite{Morant2006}, one can expect that our results for
graphene give a good indication of the outcome of actual experiments.

As a conclusion, our computational work combining classical MD
simulations and DFT total energy calculations indicate, similar to
what has been shown for carbon nanotubes, that ion irradiation is a
promising method for creating hybrid C-B/N structures for future
applications in the realm of nanoelectronics. The beneficial
irradiation energy proved out to be 50~eV with substitution
probabilities of 50\% for N and 40\% for B. At these energies the
probability for creating other defects is close to 30--50\%. By
lowering the implantation energy below the threshold for creating
single vacancies or by annealing in a suitable molecular environment
(such as NH$_3$), the vacancies created while doping can be eliminated
via dopant atoms to reach a significantly higher substitution/defect
ratio. We also reported a surprisingly effective chemical sputtering
method for N irradiation at low energies (125~eV) which leads to
production of single vacancies.
  
\section*{Acknowledgments}

We thank the Finnish IT Center for Science for generous grants of
computer time, and Academy of Finland for funding through several
projects.

\end{document}